\documentclass[sigconf]{acmart}

\copyrightyear{2019}
\acmYear{2019}
\setcopyright{iw3c2w3}
\acmConference[WWW '19 Companion]{Companion Proceedings of the 2019 World Wide Web Conference}{May 13--17, 2019}{San Francisco, CA, USA}
\acmBooktitle{Companion Proceedings of the 2019 World Wide Web Conference (WWW '19 Companion), May 13--17, 2019, San Francisco, CA, USA}
\acmPrice{}
\acmDOI{10.1145/3308560.3316455}
\acmISBN{978-1-4503-6675-5/19/05}

\usepackage{booktabs} 
\usepackage{tabulary}
\usepackage{microtype}
\usepackage{multirow}
\usepackage{tabulary}

\begin{document}
\title{Characterization of Local Attitudes Toward Immigration \\ Using Social Media}

\author{Yerka Freire-Vidal}
\affiliation{%
  \institution{Social Complexity Research Center \\ Universidad del Desarrollo}
  \city{Santiago, Chile}
}
\email{yfreirev@udd.cl}

\author{Eduardo Graells-Garrido}
\affiliation{%
  \institution{Data Science Institute \\ Universidad del Desarrollo}
  \city{Santiago, Chile}
}
\email{egraells@udd.cl}

\begin{abstract}
Migration is a worldwide phenomenon that may generate different reactions in the population. Attitudes vary from those that support multiculturalism and communion between locals and foreigners, to contempt and hatred toward immigrants.
Since anti-immigration attitudes are often materialized in acts of violence and discrimination, it is important to identify factors that characterize these attitudes.
However, doing so is expensive and impractical, as traditional methods require enormous efforts to collect data.
In this paper, we propose to leverage Twitter to characterize local attitudes toward immigration, with a case study on Chile, where immigrant population has drastically increased in recent years. 
Using semi-supervised topic modeling, we situated 49K users into a spectrum ranging from in-favor to against immigration. We characterized both sides of the spectrum in two aspects: the emotions and lexical categories relevant for each attitude, and the discussion network structure.
We found that the discussion is mostly driven by Haitian immigration; that there are temporal trends in tendency and polarity of discussion; and that assortative behavior on the network differs with respect to attitude. 
These insights may inform policy makers on how people feel with respect to migration, with potential implications on communication of policy and the design of interventions to improve inter-group relations.
\end{abstract}

\maketitle

\keywords{Immigration; Public attitude; Twitter; Semi-supervised Topic model}

\section{Introduction}
Migration is a phenomenon faced by many countries, which brings a variety of effects; both in the population from which it emigrates and in the receiving population. One of the effects that worries many countries is intolerance and hostile attitudes toward immigrants. These attitudes have been the focus of many research studies, some of which are focused on individual-level psychological and socio-economic factors \cite{burns2000economic,scheve2001labor}, and others on the contact between immigrant population and locals \cite{brown2003teammates,hopkins2010politicized,jolly2014xenophobia}. The main methods used in these studies are based on context specific surveys, which makes replication in others societies or countries difficult. 
The theories that explain the type of attitudes of locals interacting with immigrants can be summarized in two: the Intergroup Contact Theory~\cite{allport1954nature}, and the Integrated Threat Theory \cite{stephan2013integrated,nelson2009handbook}. 
The former states that people support multiculturalism and integration. 
The latter, that people think that immigrants will bring negative effects for their society, including competition for jobs and public services, worsening of the national economy, increase in crime, and the arrival of diseases. 
Particularly, the attitudes explained by the threat theory can lead to acts of violence, discrimination, and abuse; thus, it is important to understand what factors enhance such attitudes.

However, measuring attitudes is costly and impractical under dynamic scenarios. The most frequent methods are surveys, which are difficult and costly to implement. 
In this paper, we propose to make use of the information that people publish in Twitter as a proxy of their attitudes toward immigration.
It is common to find reactions and attitudes through posts in these platforms, where people express their ideas and opinions voluntarily.
We propose to define a spectrum of attitudes based on the two aforementioned theories, and to classify users and tweets into that spectrum.
We do so with a semi-supervised topic modeling technique named Topic-Supervised Non-Negative Matrix Factorization \cite{macmillan2017topic}.
TS-NMF works in a semi-supervised way because some users can be labeled as belonging to each extreme of the spectrum, something that we do with custom-built lexicons for each theory.

We perform a descriptive case study on the Chilean society, because Chile is one of the countries in which migration has reached unprecedented volume in recent years. The statistics show that immigrant population has increased from 0.8\% in 1992 to 4.35\% in 2017; and where 66.7\% of immigrants declare to have arrived mainly in 2016~\cite{censo-2017}.
For this, Chileans have developed diverse perceptions regarding the number of immigrants in the country and the phenomenon itself. To measure them with our proposed method, we collected more than 206K tweets that discuss immigration in Chile, written by more than 49K users during the year 2017. After inferring user and tweets positions in the spectrum, we performed lexical and network analysis with respect to the spectrum position.
In the lexical analysis, we used the ``Linguistic Inquiry and Word Count'' (LIWC)  lexicon \cite{pennebaker2001linguistic}, typically employed to characterize cognitive and emotional differences in discourse~\cite{harman2014measuring, coppersmith2014quantifying, gonzalez2011identifying}. 
To analyze the network structure, we estimated the polarization of the retweet and mention networks between users.

As main results, we observed that most of the discussion toward migration in Chile is targeted at Haitian migration, even though other countries have a larger share of the population. We found lexical differences in how each attitude discussed migration, and those differences were consistent with theories. For instance, social-related words were correlated with empathetic attitudes, job- and money-related words were correlated with threatening attitudes. In the network, the retweet network was polarized, in coherence as predicted by other studies regarding political discussion \cite{conover2011political,graells2015finding}. Finally, we notice that the amount and tendency of the tweets (the latter reflects the attitude towards immigration) seems to be influenced by relevant news events on national migration issues.
These results can inform public policy designers to improve inter-group relations in the country, as well as increasing the understanding of how people feel regarding an important aspect of globalization.

In summary, the contribution of this paper is two-fold. We proposed a methodology to characterize local attitudes toward migration from tweets. Then, we performed a descriptive case study in Chile using this method, obtaining results that are coherent with social theory, with added depth based on the rich information that can be extracted from Twitter.

This paper is structured as follows. Section \ref{sec:related_work} discusses the related work. Section \ref{sec:social_theories} describes the social theories that guided our analysis. Section \ref{sec:dataset} describes the data set we analyzed. Section \ref{sec:methodology} describes the methodology. Section \ref{sec:case_study} describes the results of applying the methodology to the data set. Section \ref{sec:discussion} discusses the implications of our work. Finally, Section \ref{sec:conclusions} states our conclusions.

\section{Related Work}
\label{sec:related_work}

Migration is a widely studied topic because there are many issues associated to this phenomenon. Some researchers have focused on studying the economic impacts related to migration \cite{scheve2001labor,hainmueller2007educated,hanson2007public,hainmueller2010attitudes}, others on social cohesion \cite{sniderman2000outsider,hopkins2010politicized,kopstein2009does,jolly2014xenophobia}. Within these studies, those who have focused on integration \cite{lamanna2018immigrant} and racism/xenophobia stand out \cite{brown2003teammates,herek1986instrumentality,pettigrew2006meta}. Our work seeks to contribute in the latter area, mainly due to the subject of our case study, Chile, a society that in a short time has faced a massive influx of immigrants. Migration in Chile has been a national issue, causing controversy in presidential elections, news, and municipal institutions. However, measuring attitudes is not a simple problem, nor a solved one. 
Twitter is currently a platform widely used in studies of human behavior, since it provides a valuable source of data.
Studies that have used Twitter have allowed to reveal socio-cultural characteristics of users or societies, including the level of integration of immigrants in a city~\cite{lamanna2018immigrant}, attitudes in response to triggering events, such as terrorist attacks~\cite{darwish2017predicting}, the influence of culture in personal actions~\cite{garcia2013cultural}, political polarization~\cite{garcia2013cultural}, personality traits~\cite{quercia2011our}, and personality differences between democrats and republicans~\cite{sylwester2015twitter}.

Given this body of research, we propose that Twitter can be used as a proxy to understand human behavior, in our case, the attitudes of Chileans regarding immigration.

\section{Social Theories}
\label{sec:social_theories}
The attitudes toward immigration are varied and depend on economic, socio-cultural and psychological factors. In this context, psychology and sociology have defined theories that explain the attitudes exhibited by people, who belong to different groups, when interacting with others: the Intergroup Contact Theory \cite{allport1954nature}, and the Integrated Threat Theory \cite{stephan2013integrated,nelson2009handbook}. 
The attitudes toward immigration are a particular case explained by these theories. 

\paragraph*{Intergroup Contact Theory} 
Developed in the book ``The Nature of Prejudice'' by Gordon W. Allport \cite{allport1954nature}, it postulates that prejudices are reduced when the interaction between different groups meets the following conditions: 
1) groups are on equal terms; 
2) they have common goals; 
3) there is cooperation; 
and 4) there is support from formal and/or informal institutions.
The theory states that intergroup contact reduces the fear and anxiety that exists when people interact with an unknown group \cite{stephan1985intergroup}, and that it promotes empathy and understanding towards the foreign group \cite{stephan1999role}. 

This theory has been used to ground several studies: contact between white and black people~\cite{brown2003teammates}, heterosexuals and homosexuals~\cite{herek1986instrumentality,herek1996some}, minority religious groups~\cite{novotny2011level}, and locals and immigrants~\cite{hopkins2010politicized}. 
All these studies conclude that contact improves relationships between groups.

\paragraph*{Integrated Threat Theory} 
In contrast to the Intergroup Contact Theory, the Integrated Threat Theory argues that contact between disparate groups provokes perceptions of threat and contempt~\cite{stephan2013integrated,nelson2009handbook}, for instance, due to competition for work and economic resources~\cite{ha2010consequences,esses2001immigration}. Furthermore, the threat does not have to be real, it can be subjective or fictitious~\cite{kopstein2009does}. 

This theory postulates that, when the interaction conditions are not optimal, the contact between different groups will provoke conflicting and hostile relationships. The concept of ``contact'' is not limited only to physical contact, it can also be indirect~\cite{dovidio2011improving}, imagined~\cite{crisp2009can}, and electronic~\cite{amichai2006contact,white2012dual}.

\null 

Both theories tell us what to search when we look attitudes toward immigration: attitudes motivated by empathy, in favor of immigration; and attitudes motivated by threat, against immigration. As such, we will assume that there are two attitudes, which we label as \emph{empathy} and \emph{threat}. 

\section{Data Set Description}
\label{sec:dataset}

\begin{figure}[t]
    \centering
    \includegraphics[width=\linewidth]{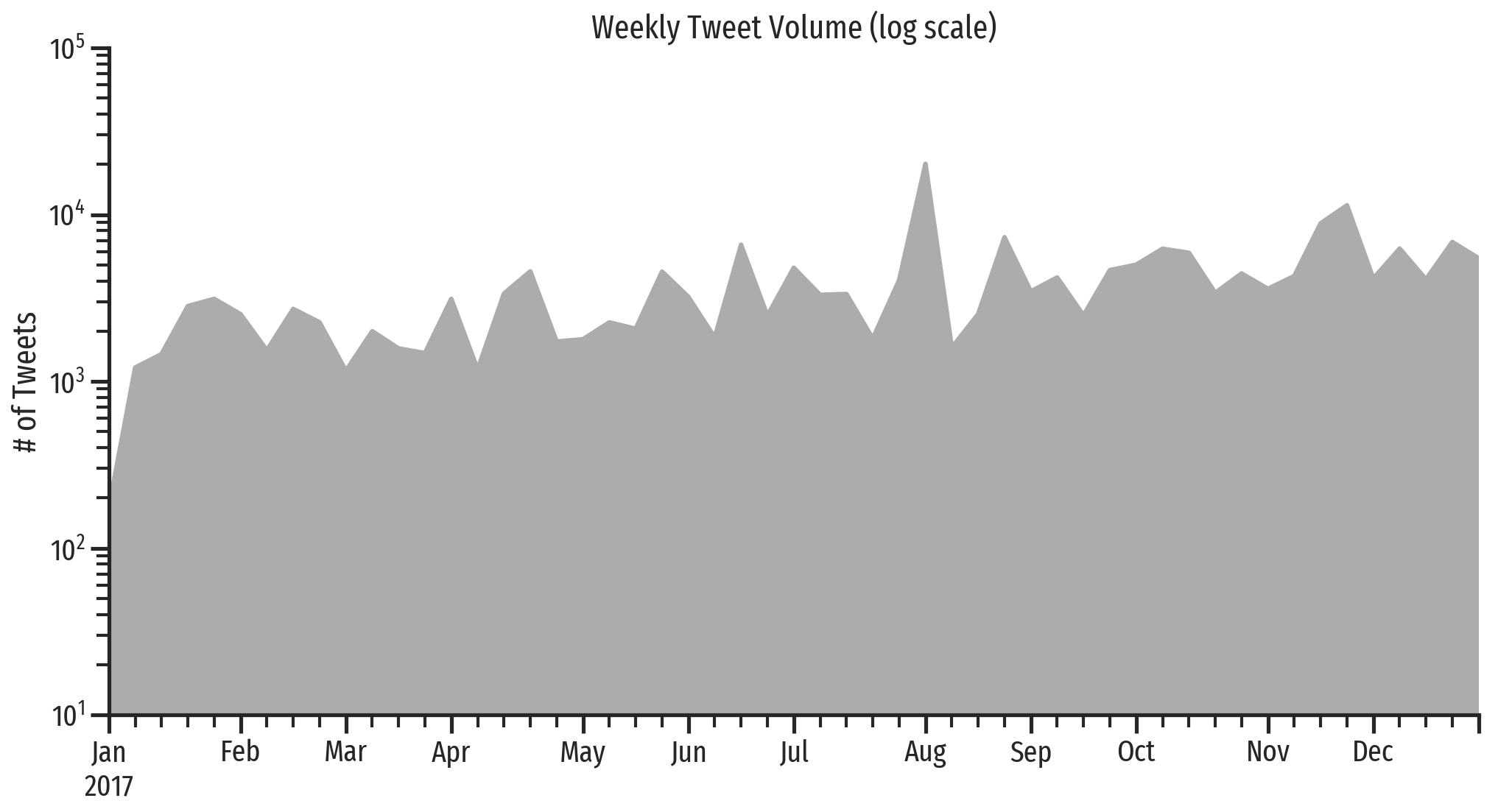}
    \caption{Weekly distribution of tweets about immigration in Chile.}
    \label{fig:dist_tweets}
\end{figure}

\begin{figure}[t]
    \centering
    \includegraphics[width=\linewidth]{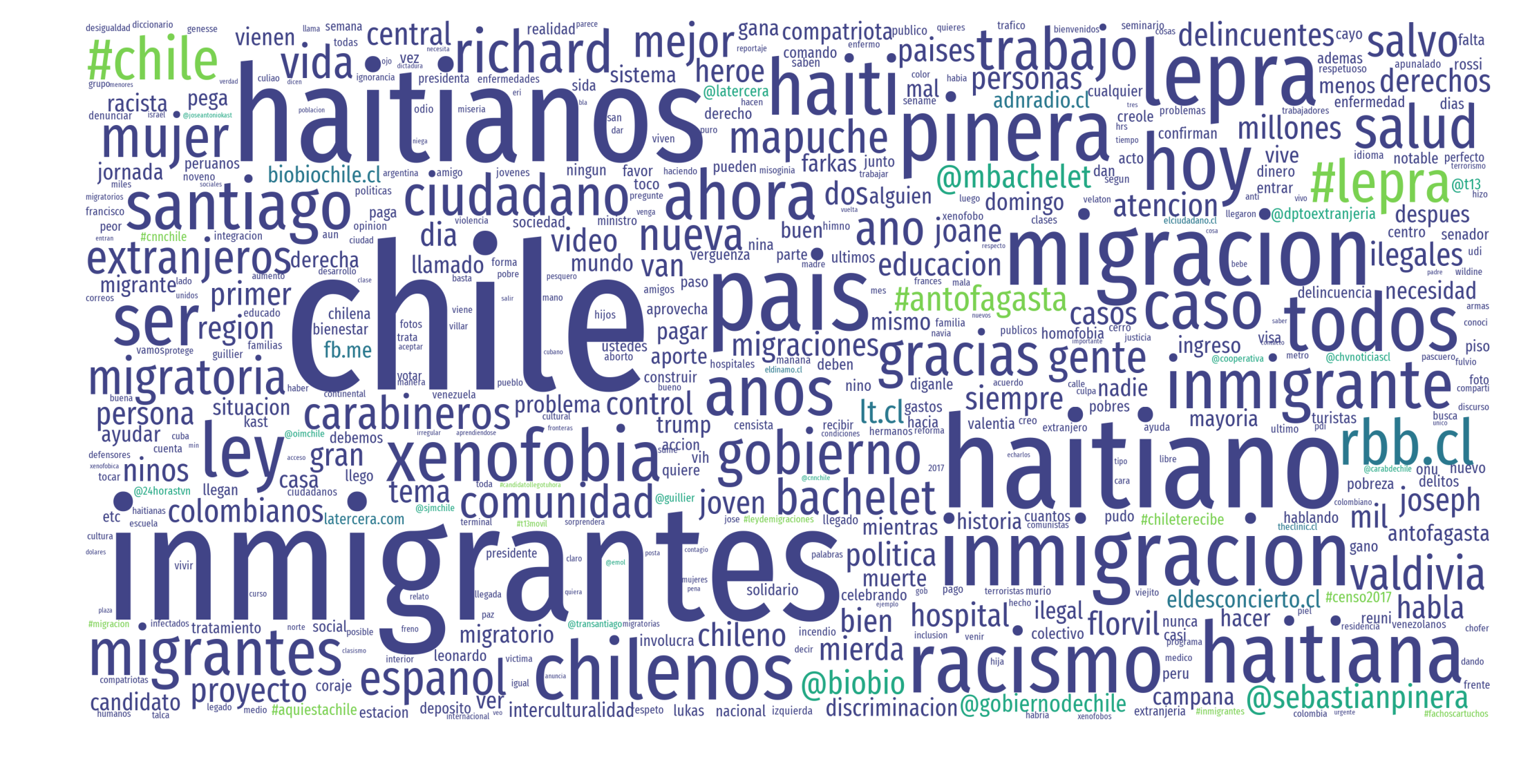}
    \caption{Wordcloud of most frequent words in the dataset, after removing stopwords. Color is assigned according to the following categories: words, hashtags, mentions, and URLs.}
    \label{fig:words}
\end{figure}

In this section we describe our data set of posts from Twitter about migration in Chile.

Twitter is a micro-blogging platform, where users publish tweets (posts) with a maximum of 280 characters.
Users may follow others, to see their tweets in their own timelines.
Tweets may mention other users, quote other tweets, or retweet another tweet to share it with one's audience.
Users can report a screen name, a full name (which can be real or fictitious), a location (real, fictitious, or empty), and a small autobiography, among other attributes.
To collect tweets that talk about immigration in Chile we used the Twitter Streaming API using system designed to crawl Chilean tweets \cite{graells2016encouraging}.
The query parameters were keywords related to immigration (\emph{e.g.}, inmigraci\'on, inmigrante, fronteras, racismo, \emph{etc.}), and origin countries with their respective demonyms (\emph{e.g.}, Hai\-t\'i--hai\-tia\-nos/as, Ve\-ne\-zue\-la--ve\-ne\-zo\-la\-nos/as, Per\'u--peruanos/as, \emph{etc.}).
Given how generic some of these keywords are, particularly regarding the context of political issues of neighbouring countries, and the presidential elections held in Chile during November and December, we performed extensive manual clean-up of the data set.

In total, our data set is comprised by 206,353 tweets that discuss immigration in Chile during 2017, written by 49,346 users.
Figure \ref{fig:dist_tweets} shows the weekly volume of tweets. As seen on the figure, the amount of tweets has a sligth positive trend. Two peaks draw our attention: July 31th, when the news reported a case of an Haitian citizen with Leprosy; and November 19th, when an Haitian citizen rescued a woman who fell from the ninth floor of a building. 

Regarding content, Figure \ref{fig:words} shows the most frequent words, after removing stopwords and accents. 
One can see that words such as Hait\'i and Haitianos are more relevant than other countries name or demonyms, despite the fact that the largest immigrant population comes from Per\'u, Colombia and Venezuela~\cite{censo-2017}. 
Also, Santiago and Antofagasta are two frequent keywords, the two cities with the largest immigrant population~\cite{censo-2017}. Other relevant words that appear are: ``gobierno'' (government), ``carabineros'' (policemen), ``Pi\~nera'' (current president, and presidential candidate in 2017) and ``proyecto'' (project); possibly because during the year an immigration reform was being discussed.

\begin{table}[t]
\caption{Examples of training terms for each attitude.}
\label{tab:table}

 \begin{tabulary}{\linewidth}{|c|C|}
  \toprule
  \textbf{Attitude} & \textbf{Training Terms} \\ \hline
    \emph{Empathy}              & \#todossomosmigrantes, \#stopxenophobia, \#chilesinbarreras, \#chileterecibe, \#bienvenidosmigrantes, @oimchile, bienvenidos a chile,  \#derribandomuros, @sjmchile, \ldots \\ 
    \midrule
    \emph{Threat}               & $\#$vendepatria, \#nomasinmigrantes, \#nomasilegales, \#inmigrantesilegales, inmigrantes delincuentes, inmigracion descontrolada, indeseables, \ldots \\
  \bottomrule
\end{tabulary}   
\end{table}

We explored the data set to seek for words, phrases, and hashtags that could be mapped to the \emph{empathy} and \emph{threat} attitudes.
In \emph{empathy} we chose terms that indicated that immigrants are welcome and will be received in equal conditions (\emph{e.g.}, ``we are all immigrants''). 
In \emph{threat} we chose terms and words that showed that immigrants are not welcome and qualified them negatively (\emph{e.g.}, ``illegal immigrants'').
Table \ref{tab:table} shows some examples of the the terms we associated to both attitudes.
These labeled terms are not necessarily frequent, however, the methodology that we describe in the next sections allows to propagate these labels through a topic model.

\section{Methodology}
\label{sec:methodology}
In this section we describe how to characterize users and tweets according to their attitude toward immigration. We define how to apply machine learning techniques to user profiles to derive user-attitude and term-attitude associations. Then, we define how to characterize attitudes from  sentiment, lexical and network perspectives.

\subsection{Attitudes and Topic Modeling}
Topic models are a family of techniques used to discover the underlying semantic structure of a corpus by identifying and quantifying the importance of representative themes in all documents
\cite{blei2012probabilistic}.
Topic models assume that each text document is generated by a set of topics which have a determined distribution. At the same time, each topic is defined by a set of words, which also have a particular distribution for each topic.

A popular topic modeling technique is Non-negative Matrix Factorization \cite{lee1999learning}.
NMF works by constructing a $k$-rank factorization of a positive document-term matrix $V$ into $W \times H$.
Matrices $W$ and $H$ are estimated by minimizing the following objective function:
\begin{equation}
    D_{NMF}\left( W, H \right)= \parallel V-  W \times H \parallel_{F}^{2}, \quad W, H \geq 0,
\end{equation}
where $\parallel \cdot \parallel_{F}$ is the Frobenius norm. 
In topic modeling, $k$, $W$ and $H$ have a special interpretation: $k$ is the number of topics, $W_{ij}$ quantifies the relevance of topic $j$ in document $i$, and $H_{ij}$ quantifies the relevance of term $j$ in topic $i$.

Typical topic modeling applications select different numbers of $k$ based on metrics such as perplexity.
However, the meaning of topics is not always interpretable, as the factorization may follow latent patterns not necessarily aligned with human expectations.
Based on the social theories described in Section \ref{sec:social_theories}, we propose to guide the learning procedure to seek for two topics: one that represents \emph{empathy}, and another that represents \emph{threat}.
In such cases, supervised methods could be employed, however, these methods require a fully labeled data set, not available in our case.
Since it is possible to map specific terms (words, phrases, hashtags, URLs, \emph{etc.}) into these two topics, we propose to use a semi-supervised version of NMF known as Topic-Supervised NMF \cite{macmillan2017topic}.
TS-NMF defines the minimization problem as follows:
\begin{equation}
    D_{TS}\left( W, H \right)=\parallel V-(W\circ L)H \parallel_{F}^{2}, \quad W, H \geq 0,
\end{equation}
where $\circ$ is the Haddamard product operator, and $L$ is a supervision matrix, defined as $L_{ij}=1$ if  topic $j$ contributes to the document $i$, and $L_{ij}=0$ if the topic $j$ does not contribute to the document $i$.
Thus, TS-NMF allows to provide examples of documents labeled with known topics, and to restrict the latent representation of the corpus to align with the labeled examples. 

In our context, we work with user profiles, \emph{i.e.}, the concatenation of tweets by a single user is one document. 
As terms we consider hashtags, mentions, URLs, and $n$-grams with $n$ up to four. This allows us to define how specific phrases are mapped to each topic.
The user corpus is represented as a document-term matrix $D$ weighted with TF-IDF \cite{baeza2011modern}, and then row-normalized with L2 norm. 
To label users in the supervision matrix, we construct a list of seed terms for each theory. 
Then, for each row in $D$ we estimate a preliminary attitude score for each topic, by adding the values of the cells of the corresponding seed terms. All users with a score above a certain threshold are labeled with the corresponding topic. In our experiments, we defined a threshold of 0.25, implying that only users who strongly used the seed terms of each topic were labeled.

As result, we obtain $D = U \times T$, where the rank of $U$ and $T$ is two. In our context, each topic is an attitude, the matrix $U$ contains the user-attitude associations, and the matrix $T$ contains the term-attitude associations (transposed). We interpret these associations as probabilities.

\subsection{Attitude Tendency and Polarity}
To characterize attitudes, we calculate two metrics common in the sentiment analysis literature  to measure the leaning and amount of sentiment: tendency and polarity \cite{kucuktunc2012large}. Tendency is defined as:
\begin{equation}
   \text{tendency}(u)=P(\text{empathy}\mid u)-P(\text{threat} \mid u), 
\end{equation}
where, $P(\text{attitude} \mid u)$ is the association between user $u$ and the corresponding attitude. Note that the definition is analog for terms. For tweets, tendency is defined as:
\begin{equation}
   \text{tendency}(\text{tweet})=\sum_{\text{term} \in \text{tweet}} \text{tendency}(\text{term}). 
\end{equation}

Note that tendency values close to zero do not imply a neutral attitude, as there could be non-zero contributions in both topics.
To clarify this fact, we consider attitude polarity as the amount of associations to both attitudes, defined for users as:
\begin{equation}
    \text{polarity}(u) = P(\text{empathy} \mid u) + P(\text{threat} \mid u).
\end{equation}
The definition for terms is analog. For tweets, polarity is defined as:
\begin{equation}
   \text{polarity}(\text{tweet})=\sum_{\text{term} \in \text{tweet}} \text{polarity}(\text{term}).
\end{equation}

In this way, tendency will allow us to group users/tweets (according to their attitude), while polarity will allow us to measure the intensity of the discussion (how polarized is the attitude).

\subsection{Lexical Characterization}
The previous metrics give an overview of user and tweet attitudes. 
The next step is to characterize grouped tweets belonging to each attitude according to their tendency.
To do so, we use a psycho-linguistic lexicon named ``Linguistic Inquiry and Word Count'' \cite{pennebaker2001linguistic}.
LIWC is a lexicon used to study emotional, cognitive and structural components contained in a text. 
In its Spanish version, it contains 7,515 words classified in one or more of 72 categories. 
Categories are classified into four dimensions: 
1) standard linguistic processes (\emph{e.g.}, articles, prepositions, pronouns, \emph{etc.}); 
2) psychological processes (\emph{e.g.}, positive and negative emotions);
3) relativity (\emph{e.g.}, time, verb tense, motion, space);
and 4) personal matters (\emph{e.g.}, sex, death, home, occupation, \emph{etc.}). 
LIWC categories are organized hierarchically, for instance, all words related to the category \emph{anger} are also organized in the categories of \emph{negative emotions} or \emph{affect} words.

We seek to estimate the association of tweets by tendency groups to LIWC categories. 
After classifying tweets into groups, we estimate how associated the words in LIWC are to each group. Note that specific events may entice a more active discussion by either group, increasing the amount of tweets, thus, we need a way to control the association with these activity patterns.
In previous work, this has been done to estimate gross community metrics with $z$-scores \cite{kramer2010unobtrusive,quercia2012tracking}. In our case, the definition is as follows:
\begin{equation}
    Z_{lt'}=\frac{P_{lt'}- \mu_{l} }{\sigma_{l}},
\end{equation}
where, $Z_{lt'}$ is the association of LIWC category $l$ with the tendency $t'$, $P_{lt'}$ is the mean of fraction of words in $l$ in each tweet with tendency $t'$, $\mu_{l}$ is the mean of fraction of words in $l$ in all tweets, and $\sigma_{l}$ is the standard deviation of the fraction of words in $l$ in all tweets. 
Hence, this relative metric allows us to compare behavior between groups, by controlling for external variability.

\subsection{Network Assortativity}
The previous definitions capture the behavior in expression, however, the social aspect of Twitter allows to also capture network behavior. We focus on two different networks: the mention network, related to discussion, and the retweet network, related to information diffusion.
In both networks, node are users, and links are weighted relations between users. Each node has as attributes its associations to each attitude.
In the mention network, a directed link between users $u_{1}$ and $u_{2}$ exists if $u_{1}$ mentions $u_{2}$ in one or more tweets. The link weight is the number of times this happens.
In the retweet network, a directed link between users $u_{1}$ and $u_{2}$ exists if $u_{1}$ republishes content by $u_{2}$. The link weight is the number of times that one user retweets another.
These kind of networks are commonly analyzed to understand polarization \cite{conover2011political}. To be able to analyze connectivity, we will focus on the Largest Strongly Connected Component of each network.

To analyze the networks structure, we estimate the assortativity coefficient with respect to each attitude. The assortativity coefficient is the Pearson correlation coefficient of numerical attributes between pairs of linked nodes (this numerical attributes are the  attitudes given by the model). It measures the similarity of connections in the graph with respect to the given numeric attribute~\cite{newman2003mixing}.
Hence, the assortativity coefficient measures whether people relations are homophilic with respect to attitude. This behavior is commonly found in networks \cite{barbera2015birds}, and it has been documented in Twitter political discussion \cite{conover2011political}, including in Chile \cite{graells2015finding}. 

\null

In the next section we apply this methodology to the data set described in Section \ref{sec:dataset}, covering an entire year of discussion about immigration in Chile.

\begin{figure}[t]
    \centering
    \includegraphics[width=\linewidth]{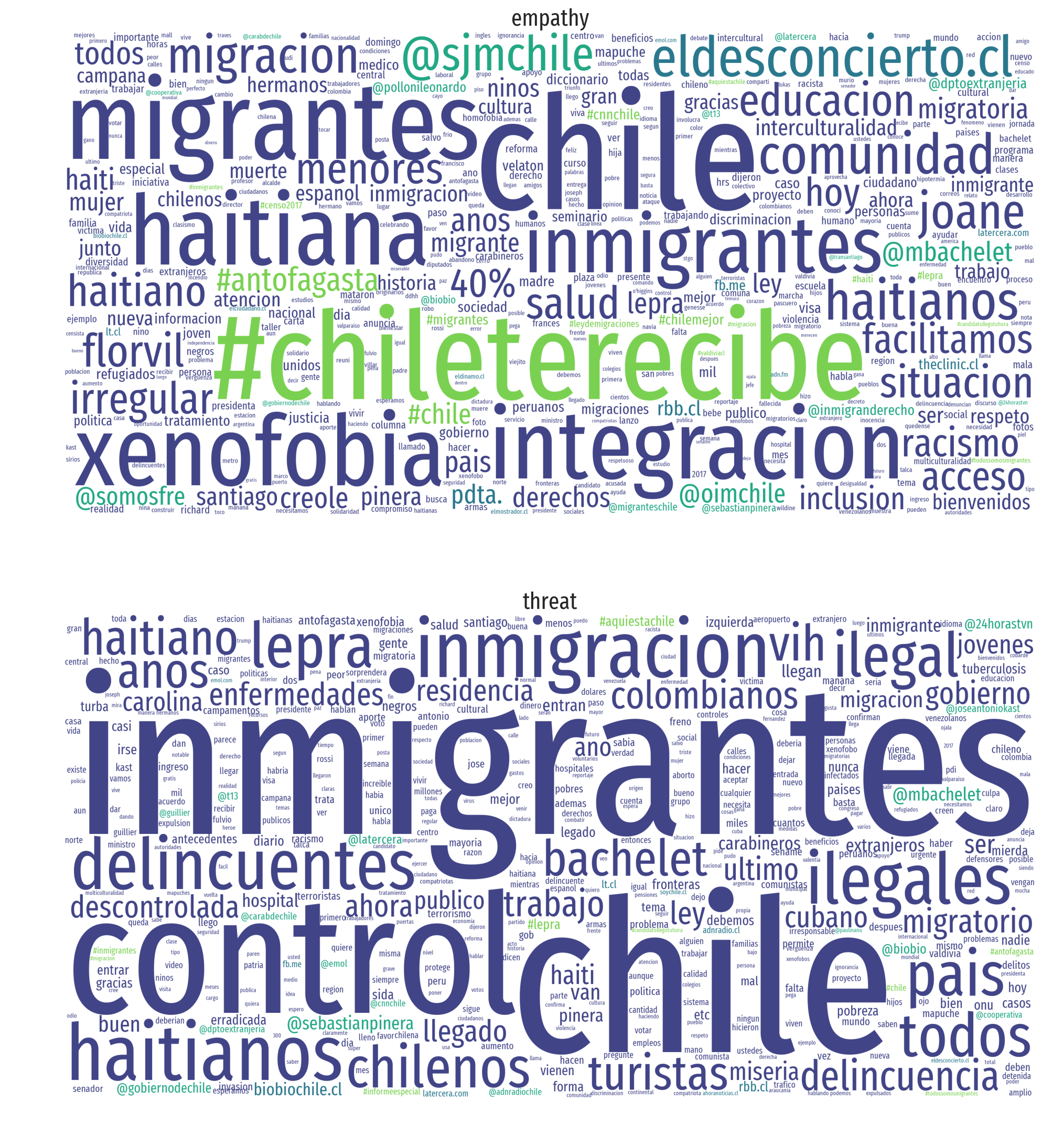}
    \caption{Most associated words to each attitude according to the TS-NMF model. Note that only single words are displayed, to avoid repetition in n-grams.}
    \label{fig:word_emp_thr}
\end{figure}

\section{Results}
\label{sec:case_study}

Here we present the results of applying the methodology from Section \ref{sec:methodology} to the data set from Section \ref{sec:dataset}.

\paragraph*{Term Associations}
Figure \ref{fig:word_emp_thr} shows the association of words with each attitude, empathy on top, threat on bottom.
One can see that words associated to empathy include ``integraci\'on'' (integration), ``salud'' (health), and ``educaci\'on'' (education), reflecting their empathetic attitude.
Words associated to threat include ``delincuentes'' (delinquents), ``control'' (control), and ``ilegales'' (illegals), reflecting a feeling of threat. Also, empathy group uses the word ``Migrantes'' (migrants) and threat group uses ``Inmigrantes'' (immigrants), which can be interpreted as that the empathy group is concerned about the general phenomenon (migration includes emigration and immigration), while the threat group only for the particular phenomenon (immigration).

\begin{figure}[t]
    \centering
    \includegraphics[width=\linewidth]{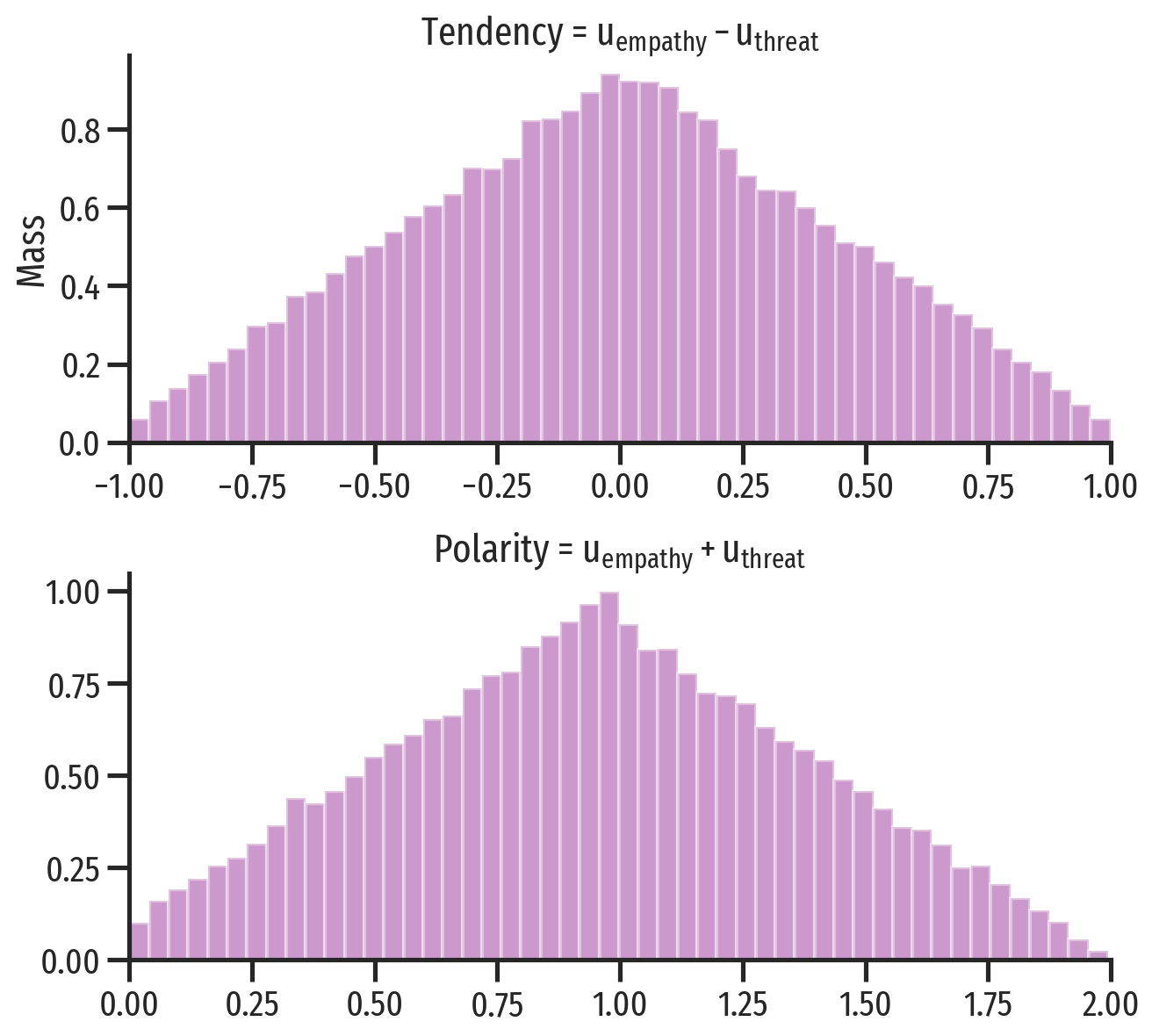}
    \caption{Top: tendency distribution for users. Bottom: polarity distribution for users.}
    \label{fig:user_tend}
\end{figure}

\begin{figure*}[t]
    \centering
    \includegraphics[width=\linewidth]{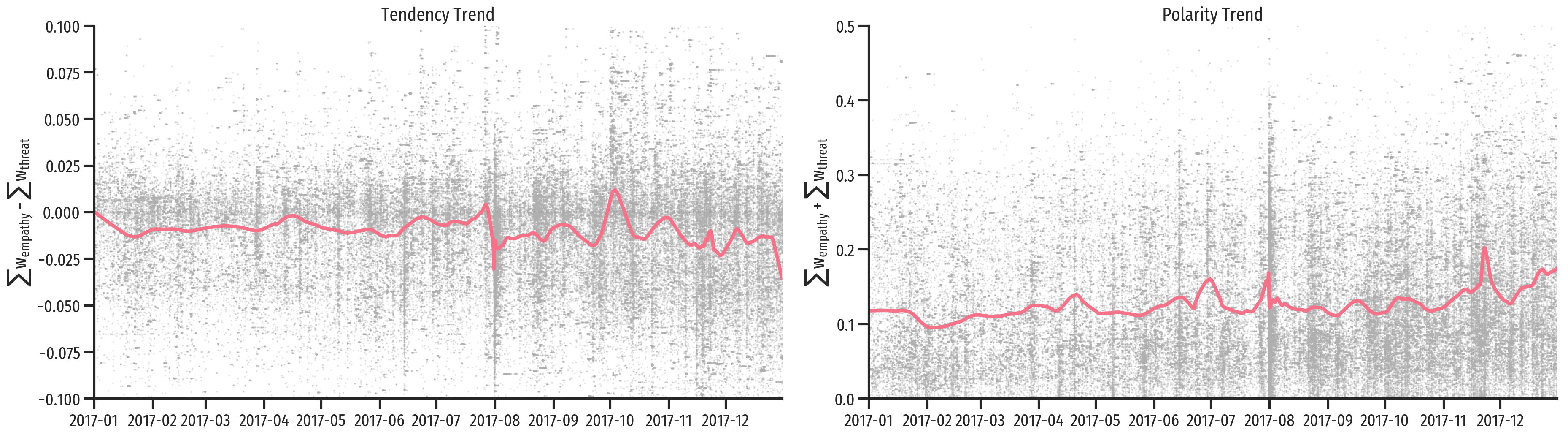}
    \caption{Trend distributions (top: tendency, bottom: polarity) for all tweets in the data set. Each tweet is a point, the $x$-position encodes its publication date, the $y$-position encodes its tendency or polarity. The line is the LOWESS interpolation of tendency and polarity.}
    \label{fig:tweets_trend}
\end{figure*}

\paragraph*{Tendency and Polarity}
Figure \ref{fig:user_tend} shows the distribution of tendency and polarity for users. One can see that the distributions are fairly symmetric, with peaks in the center of the distribution. 
Figure \ref{fig:tweets_trend} shows the tendency and polarity of tweets during the year under study, estimated using LOWESS. 
One can see that the tendency trend exhibits two interesting periods, before and after the news about the Leprosy case of an Haitian in July 31th. In the first period, tendency is slightly negative (threat), with an arguably low variability. In the second period, variability increases, and a small negative trend appears, even though at a point in time it reaches its maximum value (\emph{i.e.}, maximum empathy) at the beginning of October. %
This could be explained by a news event reported in October 6th, about a Colombian citizen that gave birth on the street because a taxi driver expelled her from his car. 

It is interesting that both news are related with the Integrated Threat theory and Intergroup Contact theory, respectively. On the one hand, the first event shows the immigrant as a threat, being a possible source of contagion of a disease (Leprosy). On the other hand, the second event shows the immigrant being a victim of violence and discrimination, which arguably makes people more empathetic.
Regarding polarity, the trend exhibits a gradual increase in time, with two interesting peaks. The first one reflects the Leprosy case, and the second one reflects the presidential elections, where migration was a common topic in discussion. 

\begin{figure*}[t]
    \centering
    \includegraphics[width=\linewidth]{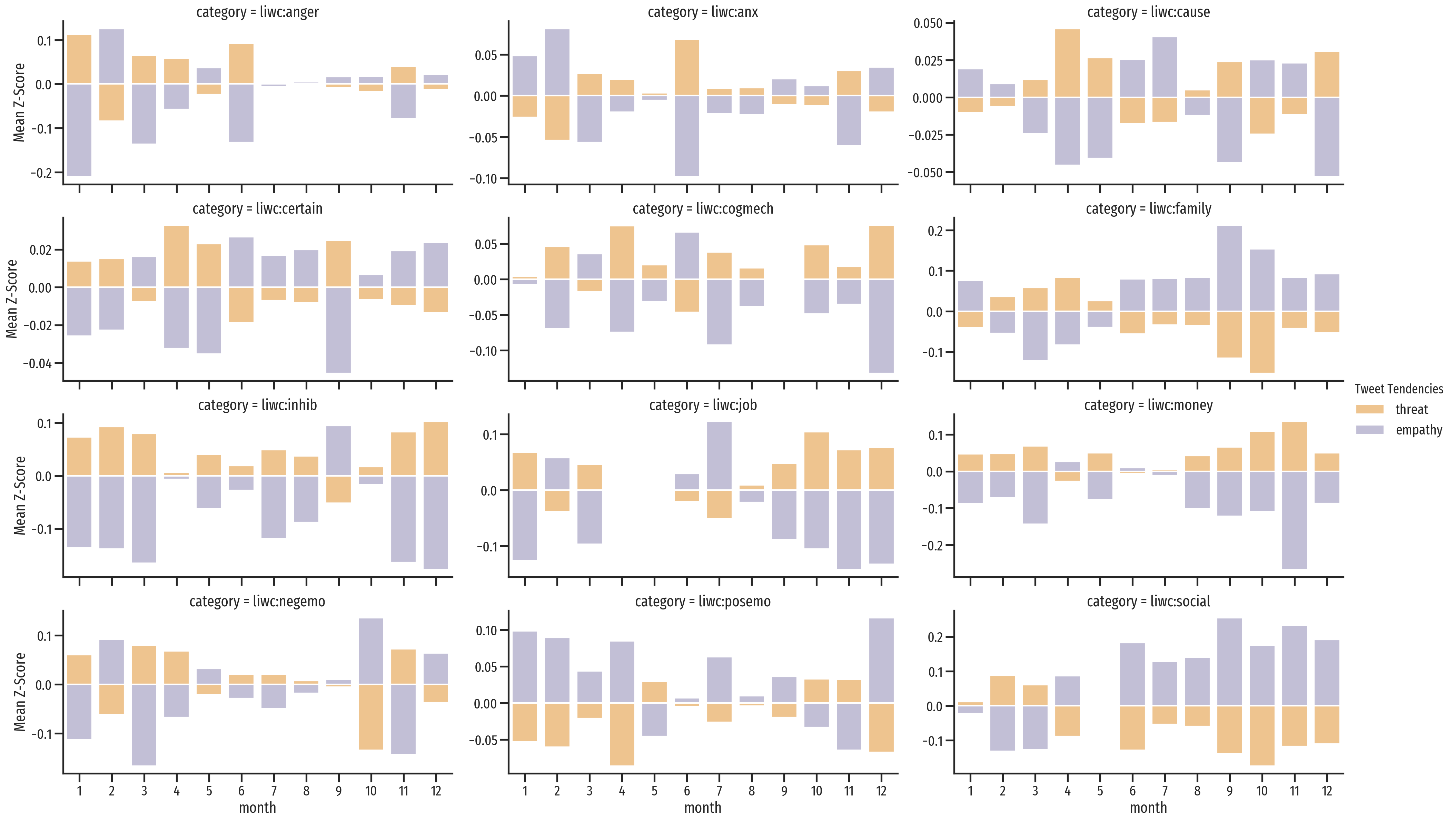}
    \caption{Association between attitudes (\emph{empathy} and \emph{threat}) and LIWC categories, per month. Each bar represents the association between groups, estimated with $z$-scores of fraction of words from each LIWC category and all other words. Purple bars indicate empathy associations, orange bars indicate threat associations.}
    \label{fig:liwc}
\end{figure*}

\paragraph*{LIWC Analysis}
Figure \ref{fig:liwc} shows the differences of cognitive and emotional categories from LIWC in tweets grouped by tendency: \emph{empathy} contains all tweets with tendency $\geq$ 0; \emph{threat}, otherwise. For each category and group, we estimated the $z$-score for all tweets each month. 
As a general observation, one can see that both groups tend to have opposite behaviors. 
For instance, tweets in the empathy group are positively associated to the \emph{sociability}, \emph{family}, and \emph{positive emotions} category more than tweets in the threat group. 
Conversely, tweets in the threat group are positively associated with \emph{money}, \emph{job}, and \emph{inhibition} categories. This could be explained by the threat theory, as immigrants can be perceived as an economic threat and labor competition. Also, inhibition category can be interpreted by the desire to prohibit the arrival of more immigrants or to prevent them from accessing social benefits.

\begin{figure*}[t]
    \centering
    \includegraphics[width=0.9\linewidth]{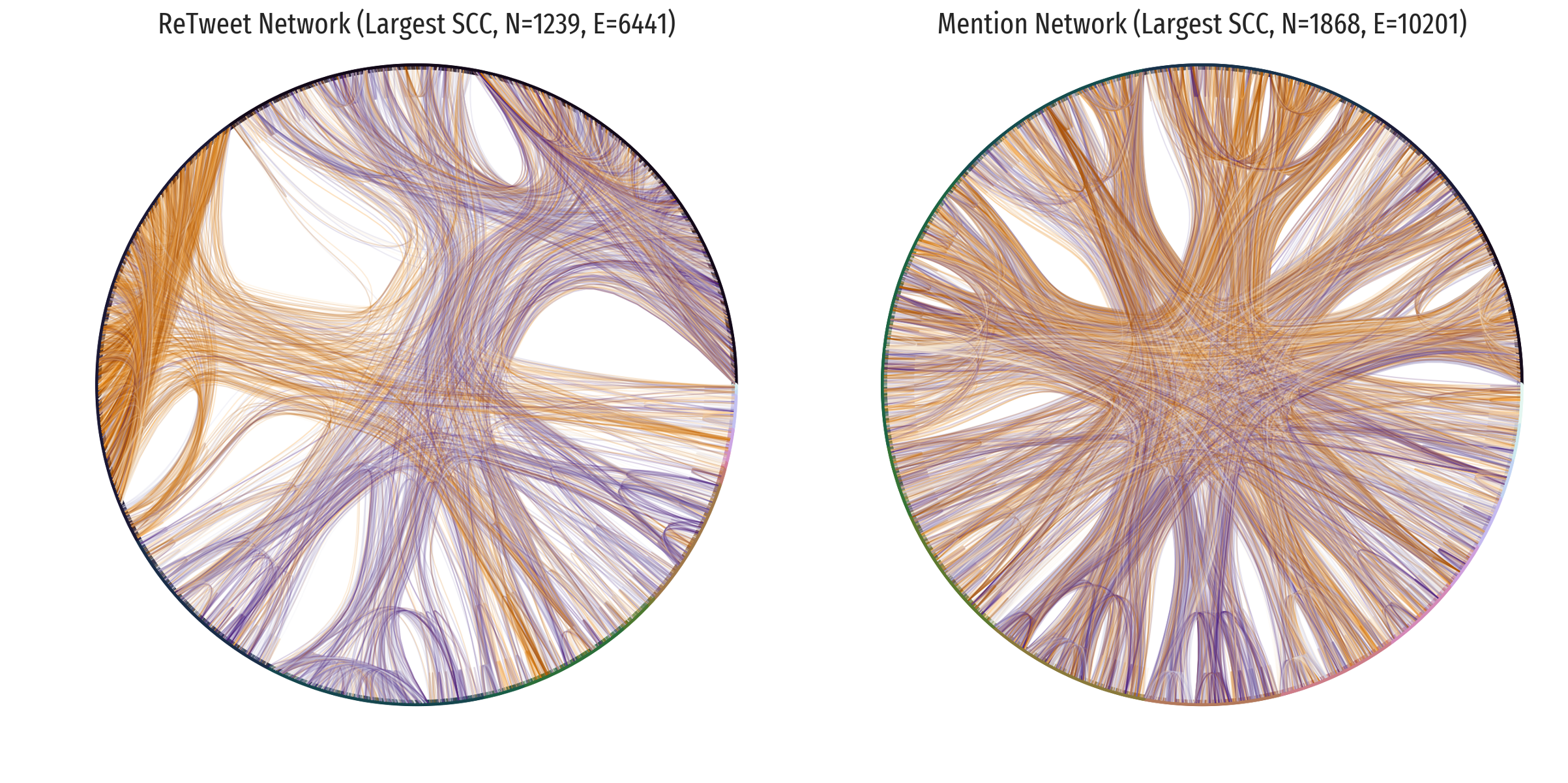}
    \caption{ReTweet Network  (left) and Mention Network (right). Each node is a circle in the outside, sorted according to the connectivity patters to other nodes. Edges are lines that join nodes, where color is the attitude of the source node (purple: empathy, orange: threat). This encoding allows to group edges that are similar in terms of connectivity between groups.}
    \label{fig:networks}
\end{figure*}

\paragraph*{Mention and Retweet Networks}
The largest SCC of the retweet network has $1,239$ nodes and $6,441$ edges, while the largest SCC of the mention network has $1,868$ nodes and $10,201$ links.
Figure~\ref{fig:networks} visualizes both networks using Hierarchical Edge Bundling \cite{holten2006hierarchical}. This method allows us to make explicit the adjacency relations between users, as similar edges are bundled to decrease visual clutter.
In the figure, each link is colored according to tendency of the source node (purple: empathy group, orange: threat group). 
Note that the visual encoding makes explicit the community structure in the retweet network and the heterogeneity of the mention network.

The assortativity coefficient for the retweet network are 0.26 (empathy) and 0.14 (threat), implying that homophilic behavior exists, but it is not as strong as in other topics (for instance, the discussion about abortion in Chile is greater \cite{graells2015finding}), and it is not equal in both groups.
As hinted by the visualization, in the mention network the results are small: 0.06 (empathy) and 0.08 (threat).
Thus, the retweet network is more segregated than the mention network. This could be explained because retweets are expected to be seen by all followers, and are a key factor in information diffusion, while mentions and replies are not. For instance, one user may send tweets to another holding an opposite position, but if there is no reply, then the interaction is not meaningful.

\section{Discussion and Future Work}
\label{sec:discussion}
Migration is a controversial issue in Chile, and, although there are some studies about Chileans attitudes toward immigration~\cite{lawrence2015crossing, carvacho2010ideological}, they do not cover recent migration patterns. 
To complement knowledge about this topic, we defined a way to classify and measure attitudes, enabling to study the dynamics of perception with respect to immigration and performed a descriptive study of how immigration is perceived in Chile, according to Twitter discussion.

Our results may inform policy and intervention design, as it quantifies how people feel and communicate with respect to immigration. This is relevant, as there exists several contact strategies to improve relationships between social groups~\cite{pettigrew2006meta}. 
For instance, the discussion we analyzed is mostly targeted at Haitian migration. A majority of them is from Afro-Haitian descent, an ethnicity that was almost non-present in Chile. 

There are two key aspects that need further exploration, and that limit the scope of our results: the representativity of Twitter, and the validation of the TS-NMF model.
In terms of representativity, Twitter is a biased sample of the population~\cite{baeza2018bias}. As such, our results only cover this sample, even though it is not know to which degree nor to which sub-populations it represents. Having these biases into account will surely improve the interpretation of results. However, one aspect that needs to be considered is that Twitter is within the most popular applications in Chile \cite{graells2018and}, and that it reflects some cultural aspects, such as the country's centralization \cite{graells2014balancing}. 
In terms of validation, the lack of ground truth or approximate measures of the problem stands in the way of effectively measuring the model accuracy, leaving us only with a qualitative evaluation.

Besides working on the limitations of our approach, there are two lines of future work that we devise.
On the one hand, it would be relevant to understand the relationship between attitudes and actual presence of immigrants in a place. This would provide a way to measure real and imagined threat attitudes \cite{kopstein2009does}. 
On the other hand, there is a potential influence of news events in attitudes. Given the rise of \emph{fake news} and \emph{post-truth} media, this would provide a way to measure the effect of such phenomena on how people feel with respect to a specific issue, migration in this case.

\section{Conclusions}
\label{sec:conclusions}
In this paper, we have characterized attitudes toward immigration by locals in Chile. We used a semi-supervised topic modeling technique (TS-NMF \cite{macmillan2017topic}) to identify attitudes grounded in two social theories, the Intergroup Contact Theory \cite{allport1954nature}, and the Integrated Threat Theory \cite{stephan2013integrated,nelson2009handbook}. Then, we measured differences in attitudes using psycho-linguistic lexicons and interaction networks. As result, we found consistent behaviour with respect to social theory. There is still work to do in the evaluation and representativeness of our model, including the definition of a suitable ground-truth perception to validate our proposal. We believe our results help to inform the design of public policy and interventions to improve relations between groups in a country. 

\bibliographystyle{ACM-Reference-Format}
\bibliography{references}

\end{document}